\documentclass[twocolumn, switch]{article} % Method A for two-column formatting

\usepackage{preprint}
\usepackage{nomencl}
\usepackage{bm}
\usepackage{braket}

\usepackage{amsmath, amsthm, amssymb, amsfonts}

\usepackage[numbers,square]{natbib}
\usepackage[utf8]{inputenc}	% allow utf-8 input
\usepackage[T1]{fontenc}	% use 8-bit T1 fonts
\usepackage{booktabs} 		% professional-quality tables
\usepackage{nicefrac}		% compact symbols for 1/2, etc.
\usepackage{microtype}		% microtypography
\usepackage{lineno}		% Line numbers
\usepackage{float}			% Allows for figures within multicol
\usepackage{newfloat}
\DeclareFloatingEnvironment[name={Supplementary Figure}]{suppfigure}
\usepackage{sidecap}
\sidecaptionvpos{figure}{c}

\usepackage{titlesec}
\titlespacing\section{0pt}{12pt plus 3pt minus 3pt}{1pt plus 1pt minus 1pt}
\titlespacing\subsection{0pt}{10pt plus 3pt minus 3pt}{1pt plus 1pt minus 1pt}
\titlespacing\subsubsection{0pt}{8pt plus 3pt minus 3pt}{1pt plus 1pt minus 1pt}

\newenvironment{conditions*}
{\par\vspace{\abovedisplayskip}\noindent
	\tabularx{\columnwidth}{>{$}l<{$} @{}>{${}}c<{{}$}@{} >{\raggedright\arraybackslash}X}}
{\endtabularx\par\vspace{\belowdisplayskip}}

\title{Analysis of Parameterized Quantum Circuits: on The Connection Between  Expressibility and Types of Quantum Gates}

\usepackage{authblk}

\author[1]{Yu Liu}
\author[2]{Kentaro Baba}
\author[2]{Kazuya Kaneko}
\author[2]{Naoyuki Takeda}
\author[1]{Junpei Koyama}
\author[1]{Koichi Kimura}

\affil[1]{Quantum Laboratory, Fujitsu Limited}
\affil[2]{Mizuho-DL Financial Technology}

\begin{document}
	
\twocolumn[ % Method A for two-column formatting
\begin{@twocolumnfalse} % Method A for two-column formatting
		
\maketitle

\begin{abstract}
Expressibility is a crucial factor of a Parameterized Quantum Circuit (PQC). In the context of Variational Quantum Algorithms (VQA) based Quantum Machine Learning (QML), a QML model composed of highly expressible PQC and sufficient number of qubits is theoretically capable of approximating any arbitrary continuous function. While much research has explored the relationship between expressibility and learning performance, as well as the number of layers in PQCs, the connection between expressibility and PQC structure has received comparatively less attention.  In this paper, we analyze the connection between expressibility and the types of quantum gates within PQCs using a Gradient Boosting Tree model and SHapley Additive exPlanations (SHAP) values. Our analysis is performed on 1,615 instances of PQC derived from 19 PQC topologies, each with 2-18 qubits and 1-5 layers. The findings of our analysis provide guidance for designing highly expressible PQCs, suggesting the integration of more RX or RY gates while maintaining a careful balance with the number of CNOT gates. Furthermore, our evaluation offers an additional evidence of expressibility saturation, as observed by previous studies. 
\end{abstract}


\vspace{0.35cm}
\end{@twocolumnfalse}]

\section{Introduction}
\label{sec:introduction}
Ever since Quantum Machine Learning (QML) was first introduced by \cite{firstQML_Lyoyd_2013}, it has been receiving increasing attention in recent years \cite{qml1,qml2}. The challenges in harnessing the quantum benefits of QML algorithms for practical applications arise from the noise and size constraints inherent in current quantum devices. One of the most promising candidates for achieving 'quantum supremacy' with the use of the 50–100 qubits in Noisy Intermediate-Scale Quantum (NISQ) devices \cite{nisq} is the QML based on Variational Quantum Algorithms (VQAs) \cite{vqa}.

In the VQA framework, computation is conducted through a synergistic approach that integrates quantum and classical mechanisms. In the quantum phase,  a carefully constructed Parameterized Quantum Circuit (PQC) generates a tunable parameterized wavefunction. This wavefunction is specifically designed to reconfigure the target problem in alignment with the encoded input states. The suitability of trial wavefunction for the target problem is evaluated by measuring the physical features of the PQC, such as the energy expectation, which serve as a cost metric. Then, guided by the cost metric, the parameters of the PQC are optimized during the classical phase to enhance the accuracy of the wavefunction. This hybrid loop of quantum and classical processes is repeated until predetermined criteria are met, such as the convergence of cost metric value or the iteration count of the loop reaching a specified threshold. Through this iterative process, the refinement of VQA’s performance is expected, ensuring its convergence towards the desired solution.
VQA-based QML models  are generally categorized into explicit quantum models and implicit kernel models \cite{qml_vqa_models2, qml_vqa_models}. This taxonomy includes quantum neural networks \cite{qnn1,qnn2}, quantum kernel methods \cite{quantum_kernel1}, and quantum data-reuploading algorithms \cite{reuploading1}. The training process for these models is carried out by tuning a selected set of parameters in the PQC.

While it remains uncertain whether QML will ultimately outperform classical machine learning algorithms for real-world applications in the NISQ era, some researchers have indicated that VQA-based QML holds the potential to provide stronger representational power than classical methods, including highly successful deep neural networks \cite{qml_vqa_advantage1,qml_vqa_advantage2,qml_vqa_advantage3,qml_vqa_advantage4,qml_vqa_advantage5}.

Since PQC holds a central position in VQA-based QML models, developing a novel PQC or selecting an established one from existing studies, such as \cite{pqc1,pqc2,pqc3,pqc4}, that fits the target problem is an essential first step in making QML model successful. 
 
Expressibility is widely adopted as a performance metric for guiding the development and selection of PQCs. It is known that a QML model is theoretically capable of approximating any arbitrary continuous function theoretically, as postulated by the universal approximation property \cite{universal_qml}. A key point of this property is that the proof assumes a high expressive PQC (along with other assumption of sufficient resources such as an ample number of qubits). Numerous efforts have been undertaken to quantify expressibility using a variety of concepts, including divergence of fidelity and Haar distribution \cite{expres1,expres2}, Fourier series transformation \cite{express_fourier}, and structure geometry \cite{exps_gnn}. 

Recent studies have investigated the relationship between expressibility and the performance of QML. In \cite{express_accuracy}, Hubregtsen et al. present their observations through a numerical analysis on 19 PQC topologies with configurations of 1 and 2 layers, revealing a strong correlation between classification accuracy and expressibility, and a weak correlation with entangling capability. These studies collectively contribute to a comprehensive understanding of expressibility as a pivotal factor in the design and optimization of PQCs for quantum machine learning applications. \cite{express_trainbility} proposes an alternating layered ansatz, a specialized hardware-efficient structure, aiming to enable the coexistence of expressibility and trainability within a PQC. Further insights into expressibility are provided in \cite{express_overfitting}, where the relationship between expressibility and overfitting in quantum learning is investigated on a hardware-efficient ansatz. 

While a lot of research has delved into exploring the relationship between expressibility and learning performance or the number of layers in PQC, there has been limited focus directed towards understanding the connection between expressibility and the types of quantum gates within PQCs. In this study, we conduct an analysis using SHapley Additive exPlanations (SHAP) values to investigate the relationship between expressibility and the types of quantum gates employed within 19 representative PQC topologies. We decompose these PQCs into elementary gates to separate the rotating and entangling functionalities. Subsequently, we generate 1,615 PQC instances by varying the number of qubits from 2 to 18 and the number of layers from 1 to 5. We then compute the KL expressibility using a quantum circuit simulator and establish a Gradient Boost Tree (GBT) model to predict Kullback-Leibler expressibility (referred to as KL expressibility in rest of the paper) with 6 types of quantum gates in PQCs. By utilizing the GBT model, we calculate the SHAP values. Finally, we evaluate the connection between expressibility and the types of quantum gates through an analysis of the SHAP values.

The remainder of this paper is structured as follows: Section \ref{preliminary} provides an overview of expressibility and SHAP values. Section \ref{pqc} outlines the specific PQCs used in our evaluation. Section \ref{expr_com} details the computation of KL expressibility. The SHAP values, computed based on a Gradient Boost Tree (GBT) model, are discussed in Section \ref{shap}. In Section \ref{relation}, we evaluate the correlation between expressibility and the types of quantum gates using the SHAP values obtained. Finally, Section \ref{conclusion} concludes our findings in this work.

\section{Preliminaries}
\label{preliminary}

\subsection{Expressibility}
Quantum computing provides a distinct advantage in efficiently processing exponentially growing data by utilizing a quantum system that expands polynomially within Hilbert space. The concept of 'expressibility' refers to the ability of a PQC to explore the Hilbert space. Numerous efforts have been made to quantify expressibility through various principles, including divergence of Haar distribution \cite{expres1}, Fourier transformation \cite{express_fourier}, and graph neural networks \cite{exps_gnn}. Among these, the expressibility, as quantified by the Kullback-Leibler divergence between the fidelity distribution of randomly sampled PQC and the Haar distribution, has gained widespread acceptance \cite{expres1,expres2,express_accuracy,express_overfitting,express_trainbility}. In this paper, we adopt Kullback-Leibler divergence-based KL expressibility as well.

In \cite{expres1}, the expressibility is defined as the Hilbert-Schmidt distance between two distributions from a state $t$-design with respect to Haar measure: the uniformly distributed  Haar distribution and the distribution generated by PQC C across the entire parameter space $\Theta$ as given in \eqref{expres_def}. Here,  $\|\cdot\|_{HS}$ denotes the square of its Hilbert-Schmidt norm, $\int_{Haar}$ denotes the integration of state $\ket{\psi}$ distributed over the Haar measure, and $\int_{\Theta}$ denotes the integration of state $\ket{\psi_\theta}$ that can be reached by PQC C within the parameter space $\Theta$.  

\begin{equation}
	\label{expres_def}
	\mathcal{A}^{(t)}(C)=\left\Vert\int_{Haar} (\ket{\psi}\bra{\psi}) ^{\otimes t}d\psi-\int_\Theta (\ket{\psi_\theta}\bra{\psi_\theta})^{\otimes t}d\theta \right\Vert_{\tiny HS}
\end{equation}    

%\if0
%\begin{equation}
%	\label{frame_potential}
%	\mathcal{F}^{(t)}(C)=\int_\Phi\int_\Theta |\braket{\psi_\phi|\psi_\theta}|^{2t}d\phi d\theta
%\end{equation}
%\fi

\begin{figure*} [pt]
	\includegraphics[keepaspectratio,scale=0.8]{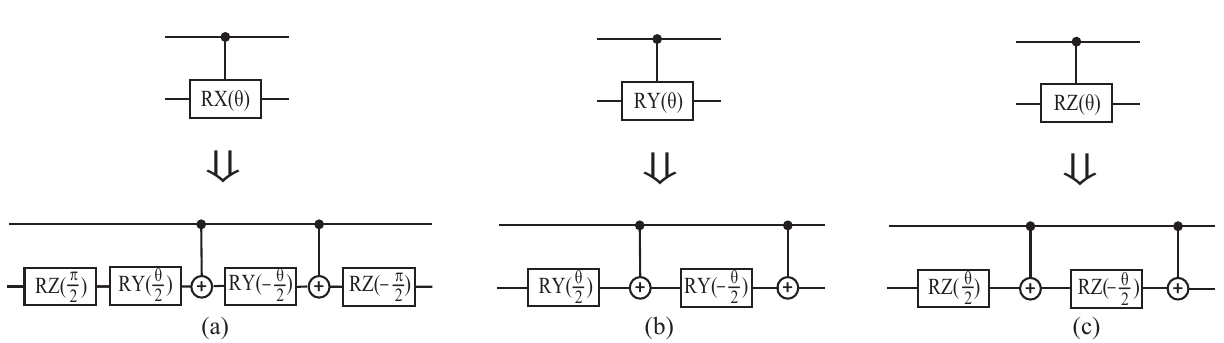}
	\centering
	\caption{Elementary decomposition of (a) CRX, (b) CRY and (c) CRZ gates}
	\label{elementary_transformations}
\end{figure*}

Following the $t$-th generalized frame potential \cite{frame_potential}, 
the expressibility can be represented as the deviation between two frame potentials,  $\mathcal{F}^{(t)}_{Haar}$ and $\mathcal{F}^{(t)}_{C}$ as shown in \eqref{expre_frame_diff}. Here $\mathcal{F}^{(t)}_{Haar}$ denotes the $t$-th frame potential with respect to Haar measure $\mathcal{F}^{(t)}_{Haar}=\int_{Haar}\int_{Haar}|\braket{\psi|\psi'}|^{2t}d\psi d\psi'$  where $N=2^n$ for $n$-qubit system, and  $\mathcal{F}^{(t)}_{C}$=$\int_\Phi\int_\Theta|\braket{\psi_\phi|\psi_\theta}|^{(2t)}d\phi d\theta$. 

\begin{equation}
	\label{expre_frame_diff}
	\mathcal{A}^{(t)}(C)=\mathcal{F}^{(t)}_{C}-\mathcal{F}^{(t)}_{Haar}
\end{equation}

The frame potential can be represented as the $t$-th moment of the distribution of fidelity $F$, where $F=|\braket{\psi_\phi|\psi_\theta}|^2 $. The deviation of frame potential, as shown in \eqref{expre_frame_diff}, is captured as KL expressibility using the  Kullback-Leibler divergence as shown in \eqref{kl_divergence}.

\begin{equation}
	\label{kl_divergence}
Expr=D_{KL}(P_C(F)\|P_{Haar}(F))
\end{equation}

The probability density function of fidelities under Haar distribution is analytically known as : $P_{Haar}(F)=(N-1)(1-F)^{N-2}$, where $N$ is dimension of Hilbert space. $P_{C}(F)$ in \eqref{kl_divergence} denotes the fidelity distribution over parameter space $\Theta$. In this study, this distribution is estimated through a numerical histogram of fidelities.  Specifically, the fidelities in the histogram are sampled by computing the fidelity of a PQC using a set of uniformly distributed $\theta$ values from the parameter space $\Theta$. Notably, a smaller value of KL expressibility indicates that the fidelity distribution is closer to Haar distribution, resulting the higher expressibility for the PQC.

\subsection{SHapley Additive exPlanations (SHAP)}

Complex learning models, e.g. gradient boost tree, deep neural networks, make it difficult to explain the connections between input feature ${\bm x}$ and prediction value $f({\bm x})$. Additive Feature Attribution Methods (AFAMs) aim to address this challenge by approximating the model’s predictions with a linear additive explain model \eqref{shap_additive} of the simplified features ${\bm x}'$, where  ${\bm x}'$ maps to the original features ${\bm x}$ by a mapping function ${\bm x}=h_x({\bm x}')$. By simplifying the original features into binary values, indicating the presence or absence of a feature, AFAMs provide a simple and intuitive way to understand how each feature contributes to the model’s prediction for a specific instance.

\begin{equation}
 \label{shap_additive}
 g({\bm x}')=\phi_0+\sum_{i=1}^M \phi_ix_i' \\
\end{equation}
where $\bm x' \in \{0,1\}^M,$ $\phi_i \in  \mathbb{R} $ and $M$ is the number of simplified features.

SHAP is proposed in \cite{shap} as one of the sophisticated and widely used AFAMs. SHAP adopts Shapley values \eqref{shapley_value} \cite{shapley} rooted from cooperative game theory to fairly distribute the contribution of input feature set $\bm x=\{x_1, x_2,\cdots, x_M\}$ to a prediction across all possible feature combinations.

\begin{equation}
	 \label{shapley_value}
	 \phi_i=\sum_{\bm S\subseteq x \backslash\{x_i\}} \frac{|\bm S|!(M-|\bm S|-1)!}{M!}[f(\bm S\cup\{x_i\})-f(\bm S)]
\end{equation}

Yong et al. have proved that Shapley value is capable of continuously satisfying three properties: local accuracy, missingness and consistency \cite{shap,shapley_prove}. Following the property of local accuracy, the prediction of input feature vector $\bm {x}^{(i)}$ for the $i$-th instance can be exactly explained by a additive model as shown in \eqref {additive_explanation}. In this model, SHAP value $\phi_j^{(i)}$ indicates the impact on prediction $\hat{f}(\bm{x}^{(i)})$ with the presence (or absence) of  $j$-th feature, and $\phi_0$ is the baseline of the prediction, e.g. $\phi_0=E(f(\bm x))$. 

\begin{equation}
	\label{additive_explanation}
	\hat{f}(\bm{x}^{(i)})=\phi_0+\sum_{j=1}^M \phi_j^{(i)}
\end{equation}

In practice, SHAP values can be computed using various methods, such as the Shapley sampling values, Kernel SHAP \cite{shap}, or Tree SHAP \cite{treeSHAP}, depending on the nature of the underlying model. Currently, the utilization of SHAP value to interpret the contributions of each input feature in complex models is prevalent across domains such as finance \cite{shap_finance1,shap_finance2}, healthcare \cite{shap_medical}, and energy \cite{shap_energy}. In this study, we employ SHAP values to analyze the impact of various quantum gates on the expressibility of PQCs.

*\begin{table*}[pt!]
	\centering
	\caption{Summary of the number of gates before and after decomposition to the elementary gates for 19 PQC topologies, each comprising 4 qubits and 1 layer.}
	\label{gate_stat}

	\begin{tabular}{c|cccccccccc}
		%\toprule
		\hline
		%\multicolumn{2}{c}{Part}                   \\
		%\cmidrule(r){1-2}
		Gate types   & RX &RY &RZ & H&CZ&CNOT&CRX&CRY&CRZ & FRZ\\
		%\midrule
		\hline
		Before &68 &44  &76 &4 &10 &14 &33&0 &33&0     \\
		After  &68 &110 &142&4 &10 &146&0 &0 &0 &66     \\
		%\bottomrule
		\hline
	\end{tabular}
\end{table*}

\section{PQCs composed of elementary quantum gates}
\label{pqc}
In this paper, we focus on analyzing the connection between expressibility and the quantum gates of various PQCs, considering factors such as PQC topology, qubit count, and layer depth. To conduct an unbiased evaluation, the set of evaluated PQCs needs to cover a broad range of topologies applied in QML, including all relevant types of quantum gates of interest.

The PQC package, as outlined in \cite{expres1} based on previous research, comprises 19 PQC topologies with 9 distinct geometries and 8 different gate types. In this work, we adopt this PQC package and further decompose the 2-qubit controlled parameterized rotation gate into 1-qubit parameterized rotation gates and 2-qubit CNOT gates, based on the decomposition method proposed in \cite{elementary_gates}, so as to separately investigate the impact on expressibility of parameterized rotation and entanglement.

Most quantum computers execute computations actually through the operations of elementary quantum gates. The decomposition into elementary gates provides a clearer representation of the circuit functionalities, facilitating deeper analysis of the connection between expressibility and various types of quantum gates.

Fig.\ref{elementary_transformations} illustrates the detailed decomposition from the controlled parameterized rotation gate to elementary parameterized rotation gates and CNOT gates. After decomposition, the combined functionalities of rotation and entanglement provided by the CRX and CRZ gates is separately provided by the RY, RZ, and CNOT gates, respectively.

Table \ref{gate_stat} summarizes the gate composition before and after decomposition to the elementary gates for 19 PQC topologies, each comprising 4 qubits and a single layer. These decomposed PQC topologies are listed in the APPENDIX. The entanglement effects introduced by CRX and CRZ operations are concentrated solely on the CNOT gate after decomposition, resulting in a significant increase from 14 to 146 in the number of CNOT gates. To decompose the CRX gate into elementary gates, 66 fixed-angle ($\pm\frac{\pi}{2}$) z-rotation gates (FRZs) are newly added. The variety of quantum gate types in the 19 PQCs has been reduced from 8 to 7, or 6 if we exclude the FRZ gate, which has no trainable parameters. Another reason for the exclusion is due to the high correlation between the FRZ gate and the RY gate, as demonstrated by a correlation coefficient of 0.99, which we will illustrate later.

\section{KL Expressibility Computation}
\label{expr_com}
By the definition in \cite{expres1}, KL expressibility can be achieved by computing the KL divergence between the pure Haar distribution $P_{Haar}(F)$ and the fidelity distribution $P_C(F)$ of PQC from numerical histogram as shown in \eqref{kl_divergence}. 

\begin{figure}[ptb]
	\centering
	\includegraphics[keepaspectratio,scale=0.5]{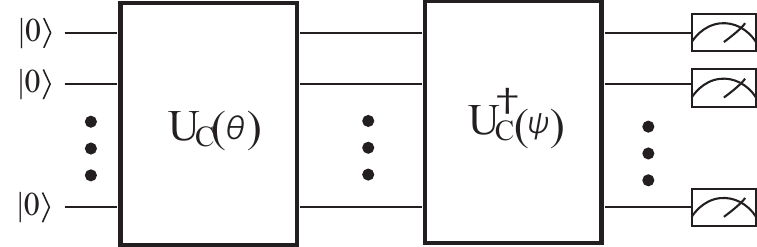}
	\caption{Fidelity computation diagram}
	\label{fidelity_diag}
\end{figure}

\begin{figure}[pt]
	\centering
	\includegraphics[keepaspectratio,scale=0.45]{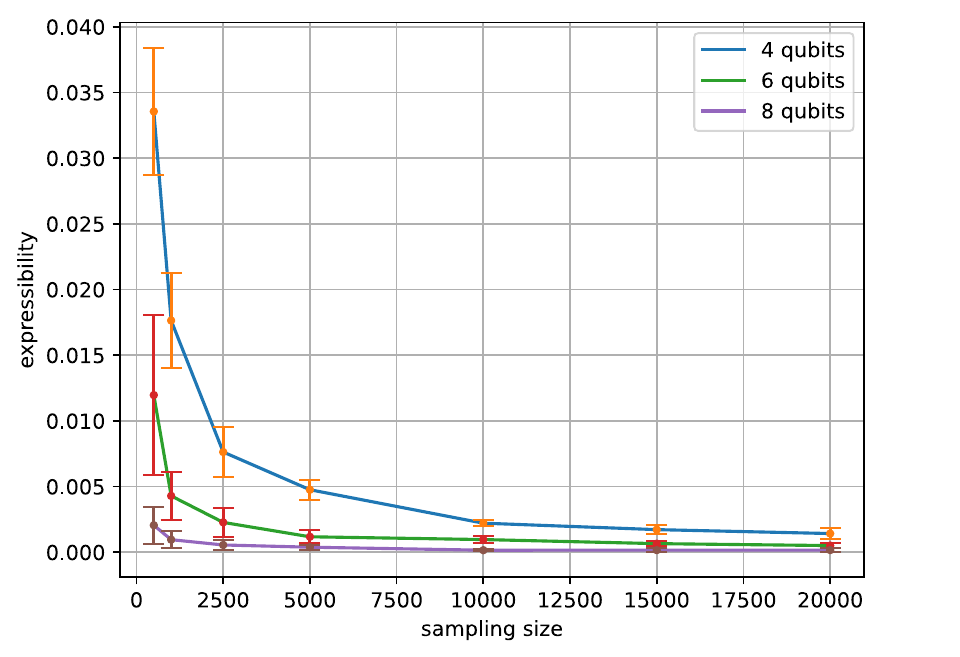}
	\caption{Variance of fidelity with increasing sampling size for the PQC with id=6, composed of 4, 6, and 8 qubits in 1 layer}
	\label{sample_std}
\end{figure}

In this work, we compute the fidelity $F=|\braket{\psi_\phi|\psi_\theta}|^2 $ of a PQC through quantum simulation of $\braket{{\mathbf 0}|U_{C}^\dagger(\phi)U_C(\theta)|{\mathbf 0}}$ with the initial state $\ket{0}^{\otimes n}$. Fig.\ref{fidelity_diag} shows the diagram of quantum circuit for the simulation, where $U_C(\theta)$ and $U_C(\psi)$ are the PQCs corresponding to parameter set $\theta,\psi$ respectively. These parameters are randomly sampled from the parameter space $\Theta$. In the simulation, the fidelity $F$ is determined as the probability of the output state $\ket{\psi}$ being identical to $\ket{0}^{\otimes n}$ for an $n$-qubit PQC. 

\begin{figure}[pt]
	\centering
	\includegraphics[keepaspectratio,scale=0.45]{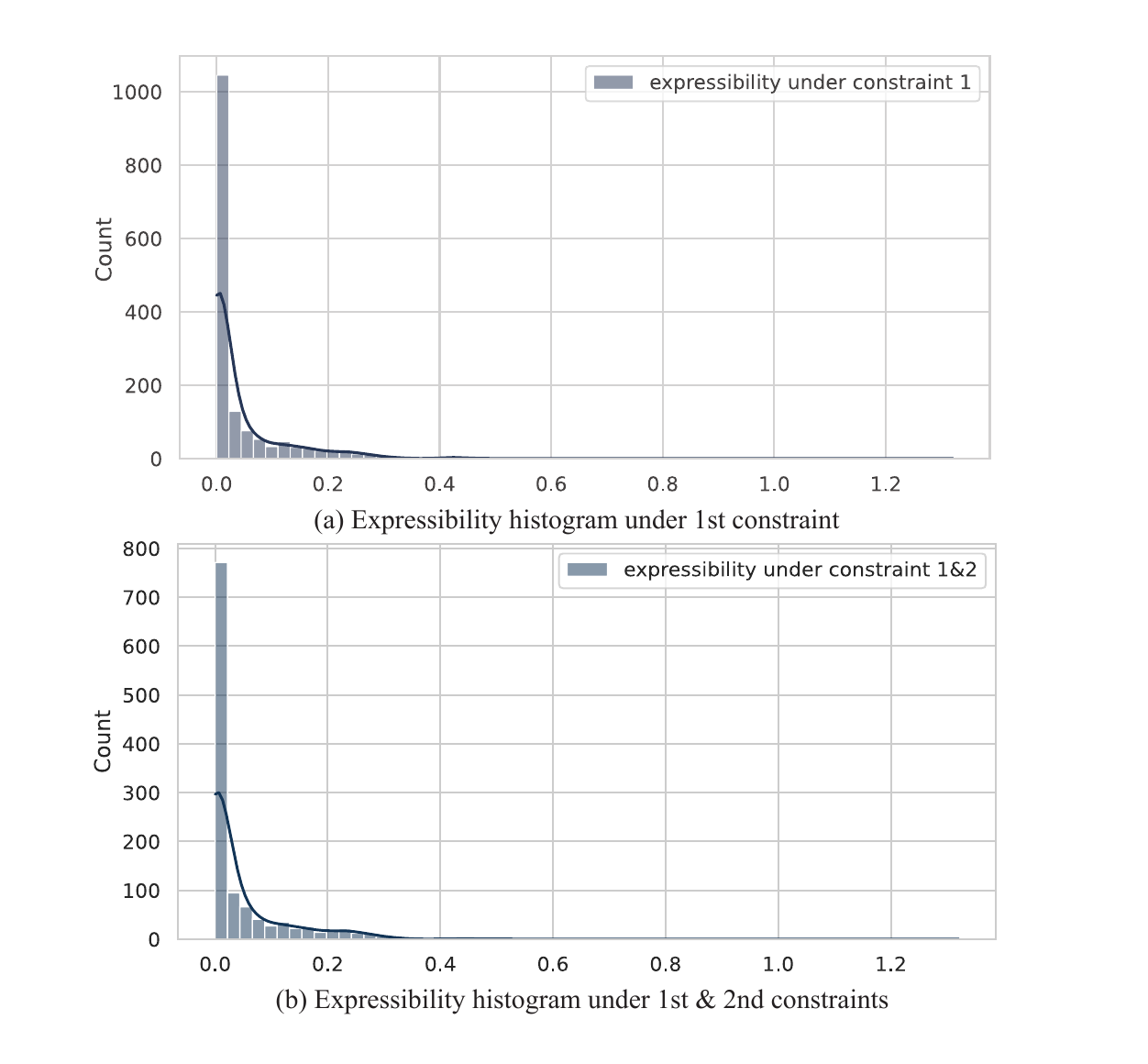}
	\caption{Histogram of KL expressibility data under the constraint conditions mentioned in the main text}
	\label{expr_hist}
\end{figure}

To mitigate the divergence bias arising from the numerical sampling of the fidelity histogram, we have set the number of sampling circuit pairs to 20,000. The final KL expressibility value is then obtained by computing the mean value across 10 iterations. Fig. \ref{sample_std} depicts the standard deviation of KL expressibility with increasing sampling size for the PQC with id=6, composed of 4, 6, and 8 qubits in a single layer. This setting appears to yield a more robust and reliable estimation of fidelity. In this work, we process the iterations in parallel to accelerate the computation.  

Our objective in this study is to evaluate the connection between expressibility and the types of quantum gates in PQCs. To achieve this, we require more variation in expressibility across different quantum gate types. Previous studies have reported the saturation of expressibility, causing expressibility to concentrate around a fixed value (typically close to zero) under specific conditions, such as increasing layers \cite{expres1} or the upper bound of Vapnik-Chervonenkis dimension \cite{express_overfitting}. To mitigate the saturation effect, we impose a constraint on the number of layers, limiting it a maximum of 5. Additionally, we individually restrict the total number of parameters to a maximum of $2^n$ because the number of parameters  is directly proportional to the number of layers. The latter restriction is also considered to maintain a quantum dimensional advantage over classical computation.

We compute expressibility data for analysis using 1,615 PQCs composed of elementary quantum gates. These PQCs are constructed by combining 19 PQC topologies, with 17 qubit configurations ranging from 2 to 18 qubits, and 5 layers ranging from 1 to 5. To assess fidelity, we conduct simulations using the state-vector-based simulator Qulacs \cite{qulacs}, allowing us to disregard errors caused by shot sampling. Fig.\ref{expr_hist} illustrates the histogram of expressibility data. We observe an further 32\% decrease in the count of expressibility values falling within the 0-bin when applying the second constraint compared to the first constraint alone.

\begin{figure}[pt]
	\centering
	\includegraphics[width=7.5cm, height=7cm]{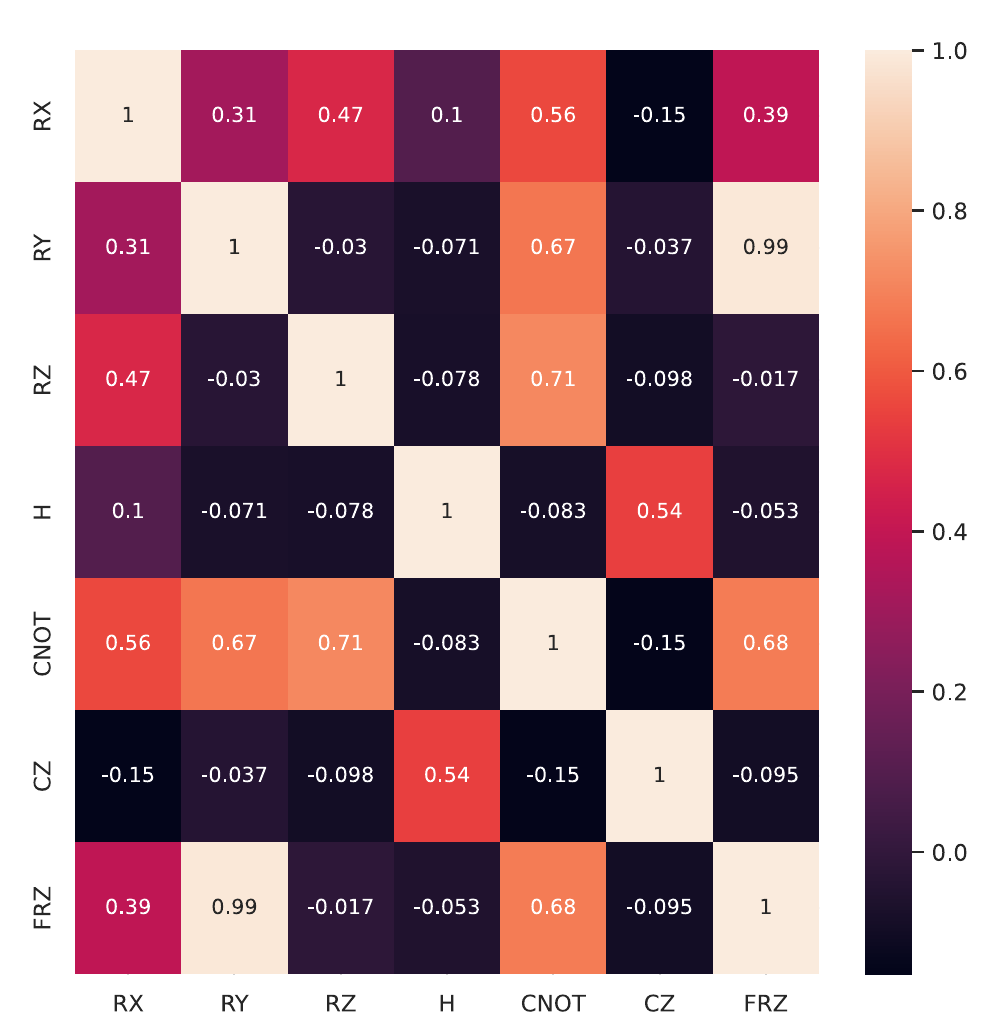}
	\caption{Pearson's correlation of gate number across different types of elementary quantum gates}
	\label{corelation_gate}
\end{figure}

\begin{figure}[pt]
	\centering
	\includegraphics[keepaspectratio,scale=0.4]{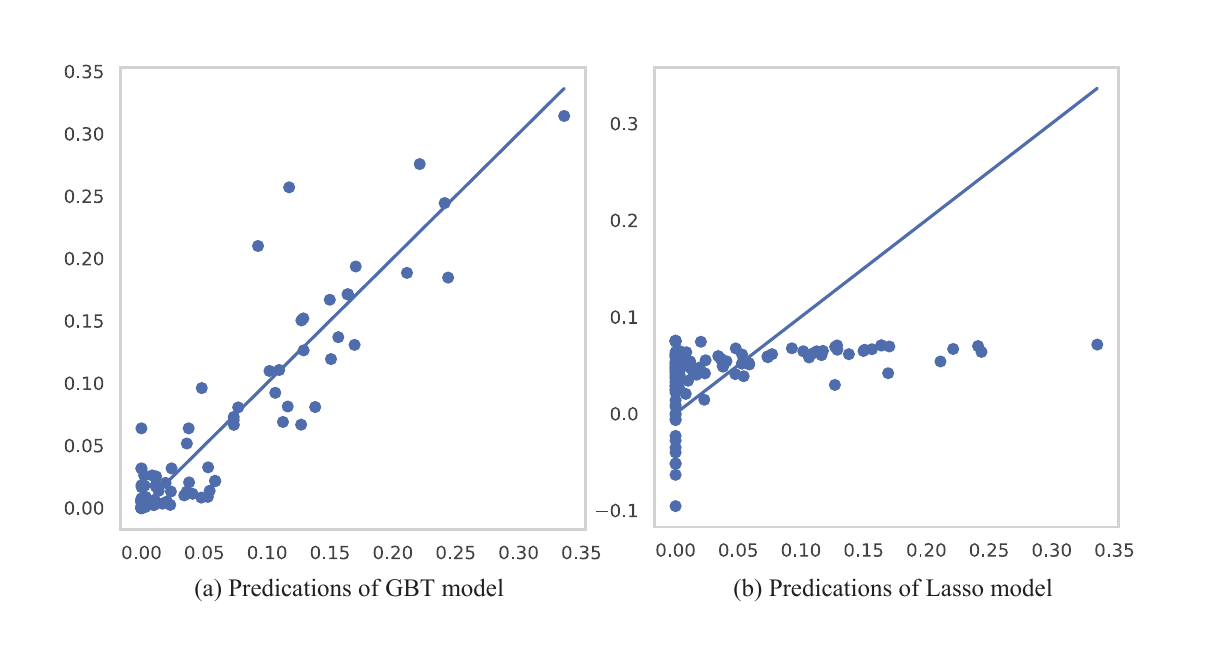}
	\caption{Prediction of KL expressibility values using (a) GBT and (b) LASSO models on the hold-out test dataset. The horizontal axis represents the true values, and the vertical axis represents the model's predicted values. The straight line indicates y=x.}
	\label{model_accuracy}
\end{figure}

\section{SHAP Values of Expressibility with the GBT Model}
\label{shap}
\subsection{Types of elementary quantum gates}
As depicted in Table \ref{gate_stat}, there are 7 distinct types of elementary quantum gates within the 19 basic PQC topologies. The correlation of gate number for 1,615 PQC instances generated based on these topologies among the 7 types of elementary quantum gates is illustrated in Fig.\ref{corelation_gate}.

From the figure, we observe that the parameterized Y-rotation gate (RY) exhibits a strong correlation with the fixed angle Z-rotation gate (FRZ), with a high correlation coefficient of 0.99. This strong correlation arises because all FRZ gates in the 1,615 PQC instances are introduced from the decomposition of the CRX gate, as the majority of RY gates are. Since we can combine the impact of the FRZ gate on expressibility with that of the RY gate through a linear combination, and the FRZ gate has no tunable parameter, we exclude the FRZ gate from this analysis. We then characterize the construction of PQCs using the normalized count of 6 types of quantum elementary gates in this study: the parameterized X-rotation (RX) gate, parameterized Y-rotation (RY) gate, parameterized Z-rotation (RZ) gate, H gate, CNOT gate, and controlled-Z (CZ) gate.

\begin{figure}[p!t]
	\centering
	\includegraphics[keepaspectratio,scale=0.4]{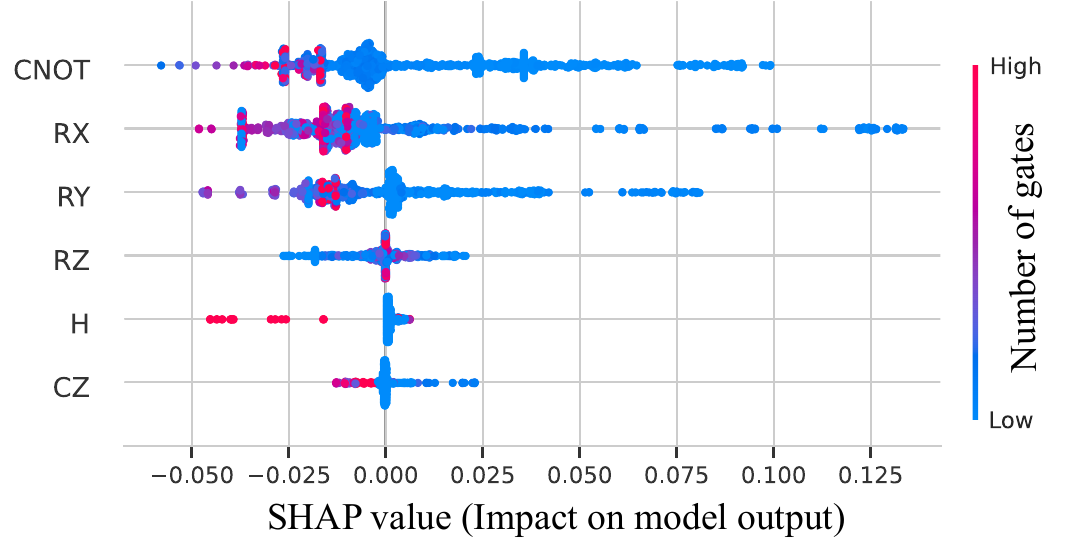}
	\caption{Beeswarm summary plot illustrating SHAP values of all the PQC instances across six elementary gates}
	\label{shap_overall}
\end{figure}

\subsection{Gradient boost tree model}

Intuitively, the linear model is one of the best models for interpreting the impact of features on an objective, assuming that the model accurately represents the objective. However, in most complex cases, the linear model exhibits poor accuracy in predicting the objective. For example, as shown in Fig.\ref{model_accuracy}, when predicting KL expressibility using the features of PQCs in this study, the linear model performs low accuracy.

\begin{figure*}[pht]
	\centering
	\includegraphics[keepaspectratio,scale=0.3]{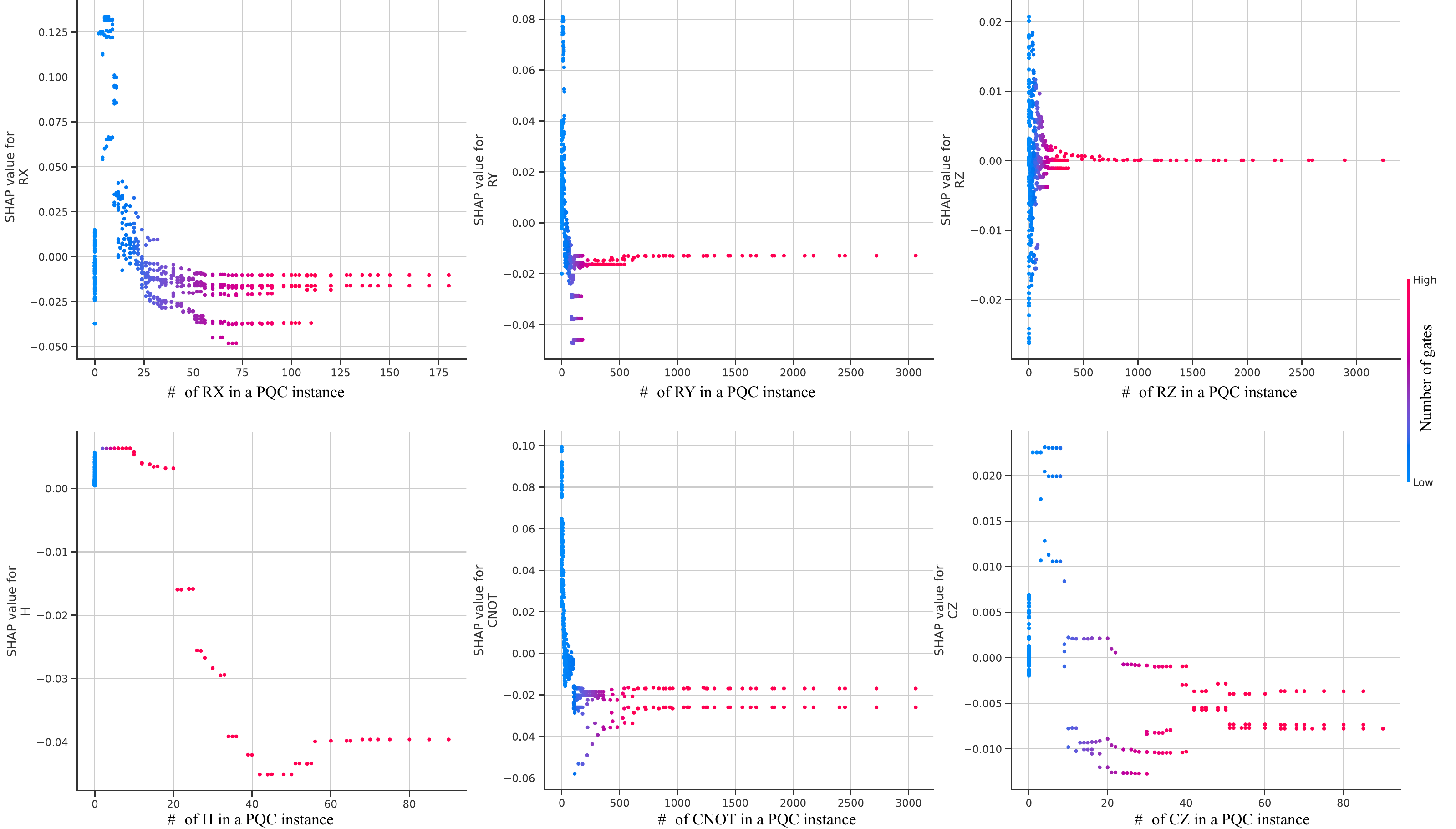}
	\caption{Dependence of SHAP values on the number of gates for each gate type}
	\label{shap_gate}
\end{figure*}

GBT models \cite{graddient_boost_tree}  are powerful tools in the machine learning, offering high predictive accuracy and robustness across various domains. The fundamental principle behind GBT models is the iterative construction of an ensemble of decision trees, where each subsequent tree corrects the errors of the previous ones. This iterative learning process, coupled with gradient descent optimization, enables GBT models to effectively capture complex patterns to  produce accurate predictions. Furthermore,  the computation of SHAP value can be significantly accelerated with using treeSHAP algorithm \cite{treeSHAP} for tree model. Among the various implementations of GBT models, LightGBM \cite{lightgbm} enhance training speed, memory efficiency and predictive accuracy compared to traditional GBT algorithms.

In this study, we employ LightGBM as the predication model of expressibility. We model the KL expressibility data over the normalized number of 6 types of elementary quantum gates that appear in the  PQCs. For the comparison, we also establish a LASSO model  using the same data. Fig.\ref{model_accuracy}  illustrates the comparison of prediction accuracy for the hold-out test dataset, which comprises 10\% of the entire dataset, randomly selected from outside the training set. We can see that the GBT model exhibits higher accuracy than the LASSO model. Specifically,  we obtain the coefficient of determination $R^2$ of 0.86 for the GBT model, and 0.21 for the linear LASSO model.

\subsection{SHAP values}

The SHAP values of all expressibility instances are then computed based on the predicative GBT model we built.   Fig.\ref{shap_overall} shows a beeswarm plot to display an information-dense summary of how various gates impact expressibility.The horizontal axis  presents the impact of each gate on expressibility, with positive values indicating weaker expressibility and lower values indicating stronger expressibility, relative to the average value $\phi_0$.  A wider range of the SHAP values indicates a greater influence  on expressbility.  Intuitively,  we observe from this summary plot that CNOT, RX and RY gates trend to have a more significant impact than RZ, H and CZ gates. 

Additionally, from Fig.\ref{shap_overall}, it can be observed that a smaller number of gates (indicated by colder colors) tends to contribute positively to the average, while a larger number (indicated by warmer colors) of gates tends to contribute negatively. This implies that if we aim for a highly expressible PQC (low expressibility value), it is worth to consider increasing the number of parameterized gates, typically by increasing the number of layers.

\section{The connection between expressibility and types of quantum gates}
\label{relation}
The SHAP values relative to the number of gates for each type of gate are depicted in Fig.\ref{shap_gate}. We reaffirm the trend that expressibility becomes stronger (indicated by smaller SHAP values in the figure) as the number of quantum gates increases, which is also depicted in Fig.\ref{shap_overall}. Furthermore, we observe that the SHAP values for all gate types eventually saturate at certain values as the number of gates increases sufficiently. This observation provides additional evidence of the expressibility saturation mentioned in \cite{expres1} and \cite{express_overfitting}, considering that the number of gates is proportional to the number of layers in the PQC. Moreover, as the number of quantum gates increases, the SHAP values reveal a decreasing or saturating trend that varies across different types of gates. This suggests that different types of quantum gates influence expressibility differently.

\begin{figure}[pht]
	\centering
	\includegraphics[keepaspectratio,scale=0.5]{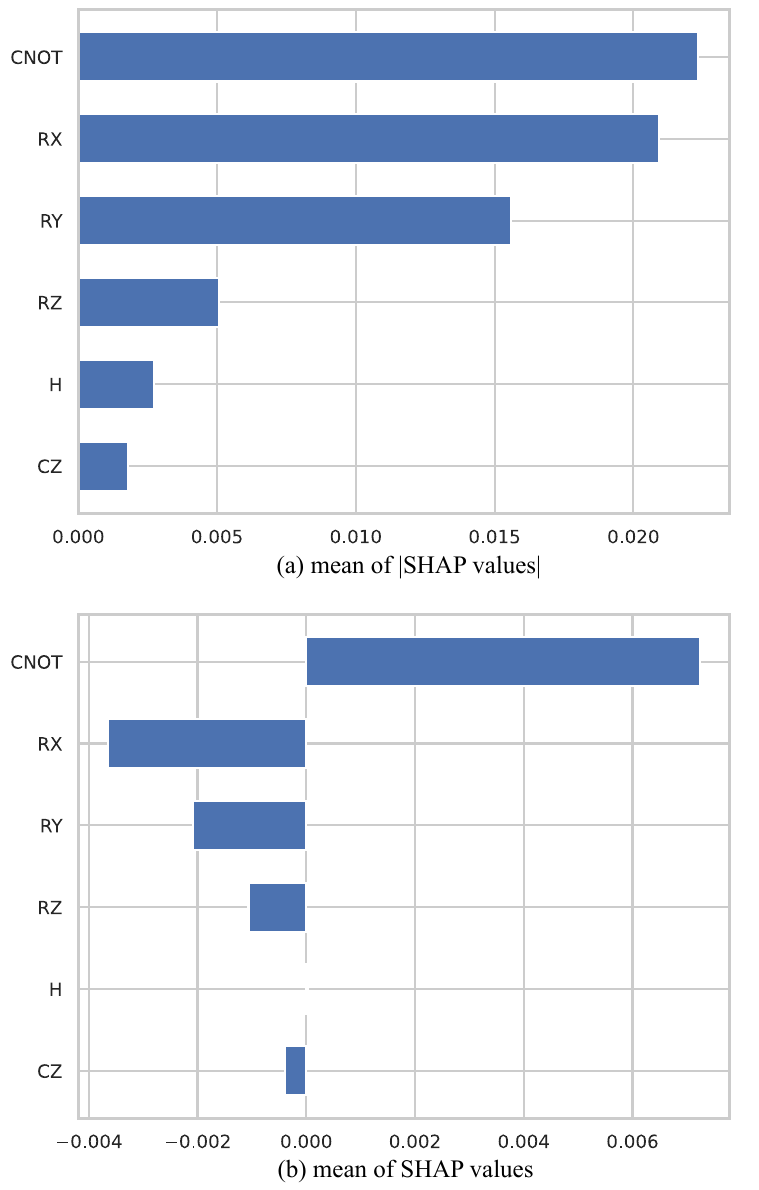}
	\caption{Average importance of quantum gate's impact on KL expressibility}
	\label{importance}
\end{figure}

Fig.\ref{importance}(a) illustrates the mean of \textit{absolute} SHAP values for each type of gate across all PQC instances, reflecting the average effect of different quantum gates on expressibility. It is evident that CNOT gate exhibits the most significant impact on expressibility, followed by the RX gate which has a slightly weaker influence. Fig. \ref{importance}(b) displays the mean SHAP values of the quantum gates. Beyond Fig. \ref{importance}(a), which presents the average absolute strength of the impact on expressibility, Fig. \ref{importance}(b) provides further perspective on the average influence on enhancing or diminishing expressibility by considering the direction of the impact. We observe that the CNOT gate tends to weaken expressibility relative to the average level, while rotational gates such as RX, RY, and RZ tend to strengthen expressibility. Among the three types of rotational gates, the RX gate appears to have the strongest effect in enhancing expressibility, followed by the RY gate. The RZ gate exhibits the weakest effect in enhancing expressibility among the three rotational gates. One reason for this is considered to be the overlap between the rotation and control axes, even though the entanglement operations are performed by the CNOT gate, as explained in \cite{expres1}.

Thus, if aiming to design or select a highly expressible PQC, it may be beneficial to integrate more RX or RY gates into the PQC by either increasing the number of layers or designing/selecting a PQC topology that contains more RX or RY gates. Simultaneously, it is  crucial to carefully balance the number of CNOT gates, taking into account other performance metrics and the complex relationships among all gates.

\section{Conclusion}
\label{conclusion}
In this paper, we analyze the relationship between the expressibility of PQCs and the types of quantum gates they encompass by employing Gradient Boosting Trees (GBT) and SHAP values. The analysis is performed on 1,615 PQC instances composed of elementary quantum gates. These PQCs are derived from 19 PQC topologies, varying in qubits ranging from 2 to 18, and layers ranging from 1 to 5. The KL expressibilities of these PQC instances are then calculated using the fidelity values computed by the Qulacs quantum simulator.

After data engineering on the expressibility dataset,  we establish a GBT model to predict expressibility based on PQC's features, the numbers of 6 types of quantum gates, achieving a coefficient of determination $R^2$ of 0.86. SHAP values are then computed through the GBT model to quantify the impact of each quantum gate type on expressibility.

The results indicate that the CNOT gate has the most significant effect on expressibility, tending to weaken it relative to the average level. Among the three rotation gates, the RX gate has the strongest impact in enhancing expressibility, followed by the RY and RZ gates. These findings provide guidance for designing or selecting highly expressible PQCs, suggesting the integration of more RX or RY gates while carefully balancing the number of CNOT gates.

%%%%%%%%%%%% Supplementary Methods %%%%%%%%%%%%
%\footnotesize
%\section*{Methods}

%%%%%%%%%%%%% Acknowledgements %%%%%%%%%%%%%
%\footnotesize
%\section*{Acknowledgements}

%%%%%%%%%%%%%%   Bibliography   %%%%%%%%%%%%%%
\normalsize
\bibliography{references}
%\renewcommand{\bibliography}{references}
%%%%%%%%%%%%  Supplementary Figures  %%%%%%%%%%%%
%\clearpage

%%%%%%%%%%%%%%%%   End   %%%%%%%%%%%%%%%%
%\end{multicols}  % Method B for two-column formatting (doesn't play well with line numbers), comment out if using method A

\onecolumn
%%\section*{Appendix}
\section*{Appendix \\4-qubit PQCs composed of elementary gates}

\begin{figure}[!h]
	\centering
	\includegraphics[keepaspectratio,scale=1]{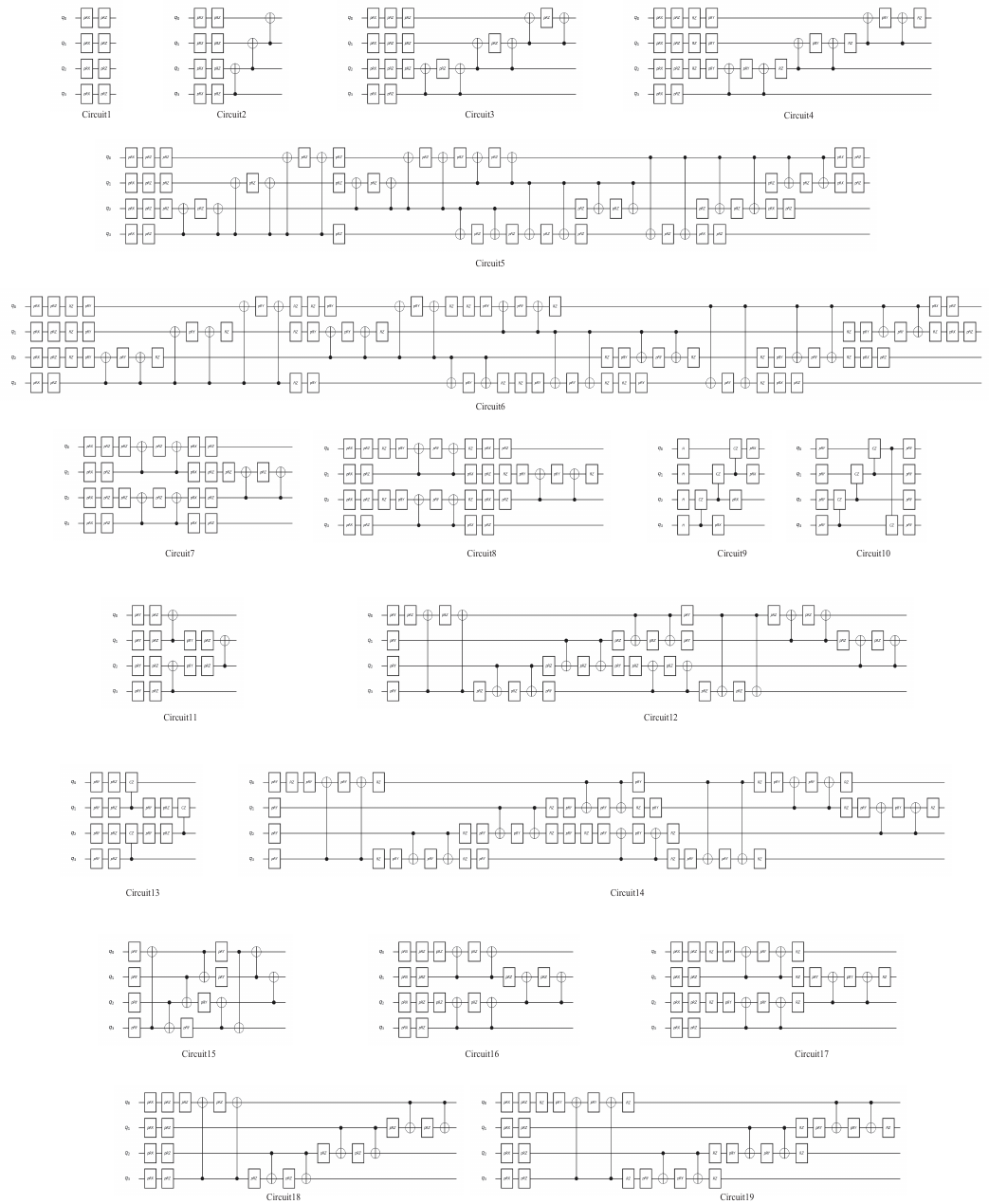}
\end{figure}

\end{document}